\documentclass[doublespacing]{elsart}

\usepackage{amssymb}
\usepackage{epsfig}
\usepackage{tabularx}
\usepackage[english]{babel}
\usepackage{graphicx}
\usepackage[numbers]{natbib}

\begin{document}
\begin{frontmatter}

\title{Low Mach number effect in simulation of high Mach number flow}


\author{X. Y. Hu} \and \author{N. A. Adams}

\address{Lehrstuhl f\"{u}r Aerodynamik, Technische Universit\"{a}t
M\"{u}nchen
\\ 85748 Garching, Germany}
\begin{abstract}
In this note, we relate the two well-known difficulties of Godunov schemes: 
the carbuncle phenomena in simulating high Mach number flow,
and the inaccurate pressure profile in simulating low Mach number flow.
We introduced two simple low-Mach-number modifications for the classical Roe flux 
to decrease the difference between the acoustic and advection contributions of the numerical dissipation.
While the first modification increases the local numerical dissipation, the second decreases it. 
The numerical tests on the double-Mach reflection problem show that 
both modifications eliminate the kinked Mach stem suffered by the original flux.
These results suggest that, other than insufficient numerical dissipation near the shock front, 
the carbuncle phenomena is strongly relevant to the non-comparable 
acoustic and advection contributions of the numerical dissipation produced by Godunov schemes 
due to the low Mach number effect.
\end{abstract}
\end{frontmatter}

\section{Introduction} \label{sec:intro}
In the last decades, Godunov schemes are among the most successful methods 
in simulating compressible flows involving shock waves and discontinuities \cite{toro2009riemann}. 
Godunov schemes utilize the solution of an Riemann solver at the face of computational cells
as the numerical flux to introduce sufficient numerical dissipation \cite{harten1983upstream}.
However, some low-dissipation Godunov schemes, such as those with Roe flux,
may suffer from numerical instabilities at the strong shock front, known as the carbuncle phenomena, 
when simulating multi-dimensional high Mach number flows \cite{quirk1994contribution}.
Although has not being fully understood,
it is generally believed that this problem is caused by 
the insufficient numerical dissipation near the shock front 
\cite{quirk1994contribution, sanders1998multidimensional, xu2001dissipative, kitamura2009evaluation, ismail2009affordable}.
Therefore, up to now, 
almost all the cures proposed for Godunov schemes is trying to increase the numerical dissipation
in the near shock-front region, or in the entire computational domain 
\cite{kim2003cures, nishikawa2008very, kim2009robust, huang2011cures}.

In this short note,
we propose to relate this problem to another well-known difficulty of Godunov schemes
in the low Mach number limit \cite{guillard1999behaviour, guillard2004behavior, dellacherie2010analysis} . 
The reason for this connection is that: 
for a shock wave in the solution of a multi-dimensional high Mach number flow,
if it propagates in one direction of the Cartesian, or nearly Cartesian grid,
the disturbances parallel to, especially behind, the front propagate in a low Mach number fashion.
Actually, the alignment of shock front to the grid is one of the typical scenarios 
of the carbuncle phenomena \cite{sanders1998multidimensional, liou2000mass, nishikawa2008very}.
\section{Roe flux and its low Mach number modifications}
For simplicity, we consider two-dimensional Euler compressible equation
\begin{equation}\label{governing-equation-2d}
\frac{\partial \mathbf{U}}{\partial t} + \frac{\partial \mathbf{F}(\mathbf{U})}{\partial x} + \frac{\partial \mathbf{G}(\mathbf{U})}{\partial y} = 0.
\end{equation}
where $\mathbf{U} = (\rho, \rho u, \rho v, E)^{T}$, $\mathbf{F}(\mathbf{U}) = [\rho u, \rho u^2 + p, \rho u v, (E + p)u]^{T}$ and $\mathbf{G}(\mathbf{U}) = [\rho v, \rho v u, \rho v^2 + p, (E + p)v]^{T}$. 
This set of equations describes the conservation laws for mass density
$\rho$, momentum density $\rho\mathbf{v} = (\rho u, \rho v)$ and
total energy density $E=\rho e +  \rho \mathbf{v}^2/2$, where $e$
is the internal energy per unit mass. 
To close this set of equations, the ideal-gas equation of state $p = (\gamma -1)\rho e$ with constant $\gamma$
is used.

The classical Roe flux gives the following form of numerical flux
\begin{equation}\label{roe}
\mathbf{F}_{i+1/2} = \frac{1}{2}\left(\mathbf{F}_{i+1} + \mathbf{F}_{i}\right) 
- \frac{1}{2}\mathbf{R}_{i+1/2}|\mathbf{\Lambda}_{i+1/2}|\mathbf{R}^{-1}_{i+1/2}(\mathbf{U}_{i+1} - \mathbf{U}_{i}),
\end{equation}
where $\mathbf{R}$ and $\mathbf{R}^{-1}$ are the right and left eigenvector matrices of $\partial \mathbf{F}/\partial \mathbf{U} $ and $\mathbf{\Lambda}$ the diagonal matrix formed with relevant eigenvalues: 
\begin{equation}\label{roe-eigenvalues}
\lambda_{1,2} = u, \quad \lambda_{3,4} = u\pm c. 
\end{equation}
The first term on the right-hand-side of Eq. (\ref{roe}) is the central flux term, and
the second term is the dissipative flux term.
Since $\mathbf{R}$ and $\mathbf{R}^{-1}$ are only forward and backward coordinate transformations,
the dissipative flux is merely dependent on $\mathbf{\Lambda}$,
and has two contributions: one is proportional to $|u|$, is called advection dissipation, 
the other proportional to $|u\pm c|$, is called acoustic dissipation.

The asymptotic analysis on Roe flux and general Godunov schemes \cite{guillard1999behaviour, guillard2004behavior} 
show that, when $M\ll 1$, where $M = |u|/c$ is the Mach number, 
or the acoustic contribution of the numerical dissipation is much larger than that of the advection,
the dissipative flux term lead to pressure fluctuations of order $\mathcal{O}(M)$ at the cell face
even if the initial data are well-prepared and contain only pressure disturbances of order $\mathcal{O}(M^2)$.
This phenomenon on inaccurate prediction of pressure profile is called low Mach number effect.
One straightforward way to decrease the low Mach number effect 
is increasing the value of $M$ in $\mathbf{\Lambda}$ 
to obtain comparable acoustic and advection contributions. 
This leads to two simple modifications of $\mathbf{\Lambda}$ to $\mathbf{\Lambda}'$ 
\begin{equation}\label{roe-eigenvalues-m1}
\lambda'_{1,2} = u, \quad \lambda'_{3,4} = u\pm \min(\phi|u|, c),
\end{equation}
and to $\mathbf{\Lambda}''$
\begin{equation}\label{roe-eigenvalues-m2}
\lambda''_{1,2} = u'', \quad \lambda''_{3,4} = u''\pm c, \quad u'' = {\rm sgn}(u)\max\left[\frac{c}{\phi}, |u|\right],
\end{equation}
where $\phi$ is a positive number of order $\mathcal{O}(1)$.
The Roe flux with Eq. (\ref{roe-eigenvalues-m1}) is denoted as Roe-M1 and with Eq. (\ref{roe-eigenvalues-m2}) as Roe-M2.
Unlike Guillard and Viozat \cite{guillard1999behaviour}  and Li et al. \cite{li2009development},
these modifications do not change the eigenvector matrix or the central flux term.
Note that, though both modifications lead to comparable 
acoustic and advection contributions of numerical dissipation
in the low Mach number region or direction of the flow,
while the Roe-M1 flux decreases the local numerical dissipation,
the Roe-M2 flux increases it.
\section{Simulation and discussion}
We test the calssical Roe, Roe-M1 and Roe-M2 fluxes, with $\phi = 5$, 
for a problem from Woodward and Colella \cite{woodward1984numerical} 
on the double Mach reflection of a Mach 10 shock in air. 
The initial conditions are
$$
(\rho, u, v, p)=\cases{(1.4, 0, 0, 1) & if $y < 1.732(x - 0.1667)$ \\
\cr (8, 7.145, -4.125, 116.8333) & else \\
},
$$
and the final time is $t = 0.2$. The computational domain of this problem is $[0,0]\times[4,1]$.
Initially, the shock extends from the point $x = 0.1667$ at the bottom
to the top of the computational domain.
Along the bottom boundary, at $y = 0$, the region from $x = 0$ to $x = 0.1667$
is always assigned post-shock conditions, whereas a reflecting wall condition
is set from $x = 0.1667$ to $x = 4$. Inflow and outflow boundary conditions
are applied at the left and right ends of the domain, respectively.
The values at the top boundary are set to describe the exact
motion of a Mach 10 shock.
The calculations are carried out with the 5th-order WENO-Z \cite{borges2008improved} scheme,
in which the Roe approximation is used for the characteristic decomposition at the cell
faces and the 3rd-order TVD Runge-Kutta scheme is used for time integration \cite{shu1988efficient}. 
Note that, in order to exclude the influence of entropy condition 
\cite{kitamura2009evaluation, ismail2009affordable}, 
no entropy fix is used in the computations.

\begin{figure}[p]
\begin{center}
\includegraphics[width=\textwidth]{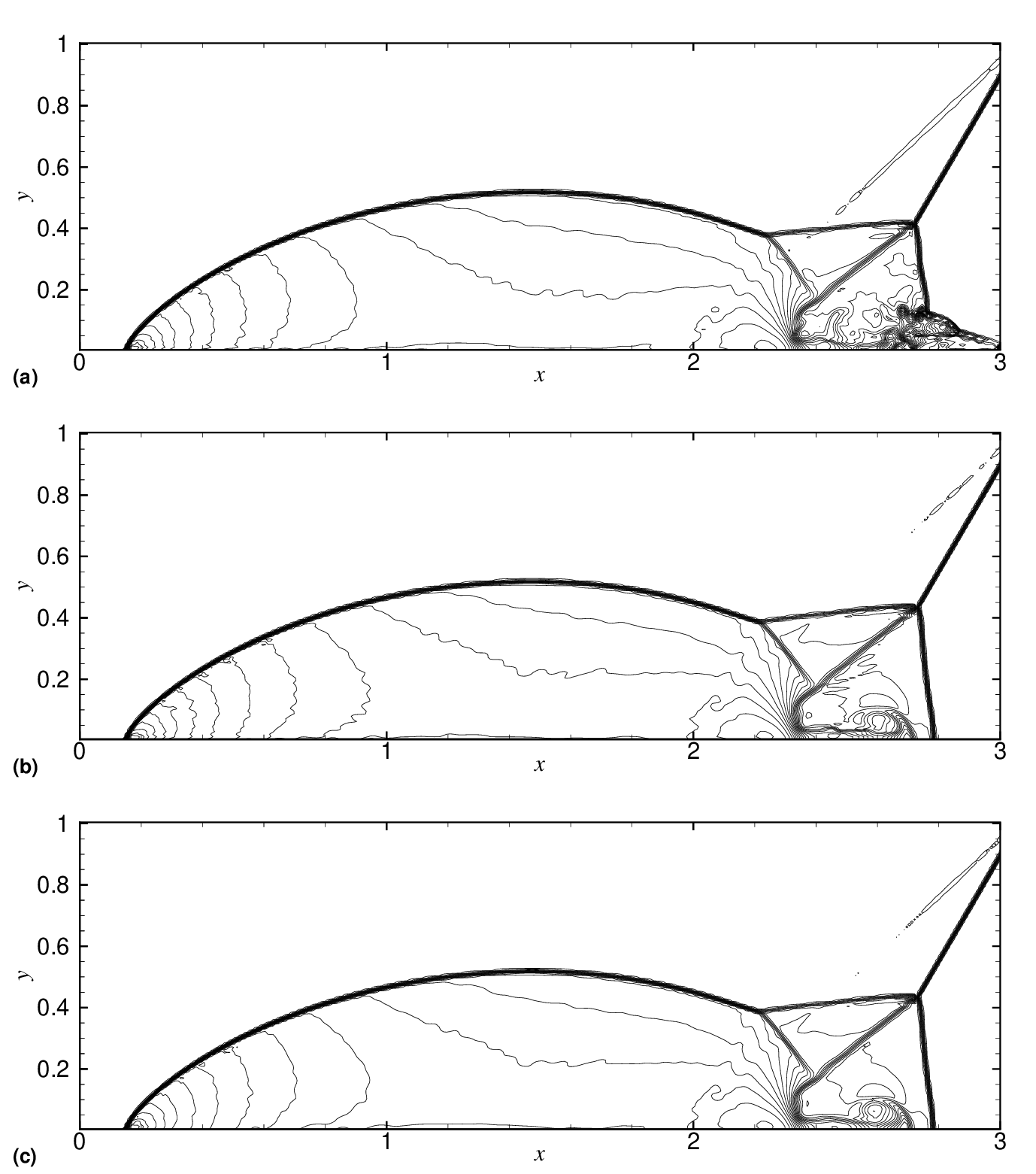}
\caption{Double-Mach reflection of a Mach 10 shock wave: 40 density contours from 1.88783 to 20.9144 at $t = 0.2$,
on a $480\times 120$ grid with the (a) classical Roe, 
(b) Roe-M1, and (c) Roe-M2 fluxes.} \label{double-mach}
\end{center}
\end{figure}

Figure \ref{double-mach} shows the density contours of the solution on a $480\times 120$ grid at $t=0.2$.
Note that, a positivity-preserving scheme \cite{Hu2012positivity} 
is implemented to obtain numerically stable results for the classical Roe flux.
It is observed that, contrast to that the classical Roe flux suffers from a kinked Mach stem, 
a typical configuration of the carbuncle phenomena,
both Roe-M1 and Roe-M2 fluxes produce numerically stable and correct results.
While Roe-M2 flux eliminates the kinked Mach stem 
by introducing extra advection dissipation parallel to the shock front,
Roe-M1 flux achieves this by decreasing acoustic dissipation.
Note that, since Roe-M1 decreases the overall local numerical dissipation near the shock front,
it is unlikely that the kinked Mach stem is produced due to 
insufficient numerical dissipation.
Further numerical experiments show that, 
if the numerical dissipation is modified by only increasing 
the value of $c$ in Eq. (\ref{roe-eigenvalues}) 
i.e. only the acoustic dissipation increased,
the simulation still suffers from the kinked Mach stem (not shown here).
All these results clearly suggest that
the carbuncle phenomena is strongly relevant to 
the non-comparable acoustic and advection contributions of the numerical dissipation produced 
by Godunov schemes due to the low Mach number effect.
\bibliographystyle{plain}
\bibliography{../reference/turbulence}
\end{document}